\documentclass[%
 reprint,
 showpacs,
 aps,
 superscriptaddress,
 pra,]{revtex4-2}

\usepackage{bm}
\usepackage{amsmath}
\usepackage{amssymb}
\usepackage{amsfonts}
\usepackage{braket}
\usepackage[dvipdfmx]{graphicx}
\usepackage{color}
\usepackage{here}

\begin{document}

\title{Preconditioning for a Variational Quantum Linear Solver}

\author{Aruto Hosaka}
\author{Koichi Yanagisawa}
\author{Shota Koshikawa}
\author{Isamu Kudo}
\author{Xiafukaiti Alifu}
\author{Tsuyoshi Yoshida}
\affiliation{%
Information Technology R\&D Center, Mitsubishi Electric Corporation, Kamakura 247-8501, Japan
}%

\date{\today}

\begin{abstract}
We apply preconditioning, which is widely used in classical solvers for linear systems $A\textbf{x}=\textbf{b}$, to the variational quantum linear solver. By utilizing incomplete LU  factorization as a preconditioner for linear equations formed by $128\times128$ random sparse matrices, we numerically demonstrate a notable reduction in the required ansatz depth, demonstrating that preconditioning is useful for quantum algorithms. This reduction in circuit depth is crucial to improving the efficiency and accuracy of Noisy Intermediate-Scale Quantum (NISQ) algorithms. Our findings suggest that combining classical computing techniques, such as preconditioning, with quantum algorithms can significantly enhance the performance of NISQ algorithms.
\end{abstract}

\maketitle

\begin{figure*}[t]
    \includegraphics[width=16.8cm]{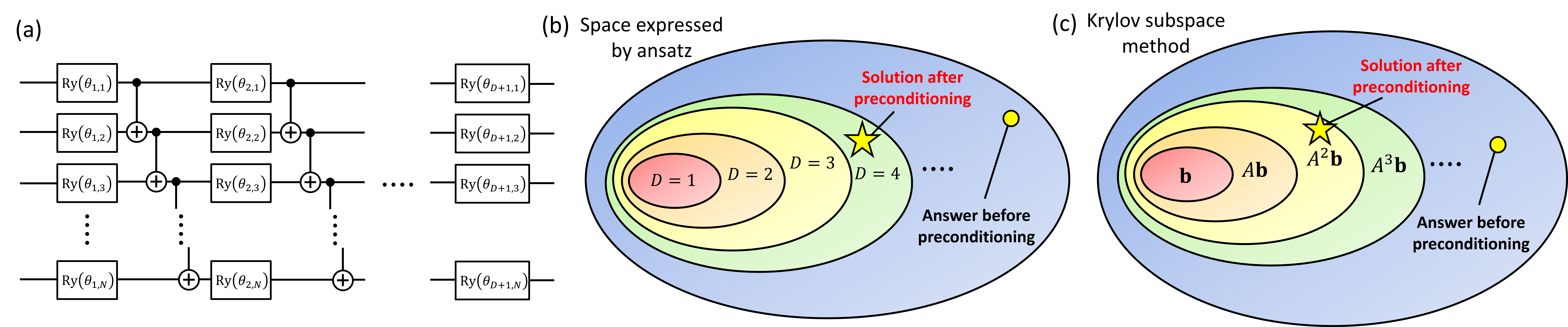} 
    \caption{(a) An example of a hardware-efficient ansatz. The ansatz consists of alternating RY gates acting on individual qubits and CNOT gates between adjacent qubits, repeated $D$ times, resulting in quantum state amplitudes being real numbers. The rotation angles $\theta$ of the RY gates can be independently determined and optimized by an optimizer. (b) and (c) illustrate the effect of preconditioning in reducing the required depth $D$ of the ansatz in VQLS, and the impact on reducing the search space in the Krylov subspace method for classical iterative linear solvers, respectively.}
    \label{fig1}
\end{figure*}

\section{Introduction}
Recently, Variational Quantum Linear Solvers (VQLS) have emerged as promising quantum algorithms for solving sparse linear systems of equations \cite{vqls1,vqls2} on Noisy Intermediate-Scale Quantum (NISQ) machines \cite{NISQ1}, with potential applications across various domains, including computational fluid dynamics simulation \cite{QFlowS,heat}, finite difference time domain method \cite{FDTD,FDTD2}, and machine learning \cite{QML,etc1,etc2}. VQLS leverages the power of NISQ machines by employing parameterized quantum circuits, known as ansatz, to represent and approximate solutions to linear equations. However, the effectiveness of VQLS depends heavily on the ability of the ansatz to express the optimal solution within its parameterized space \cite{dynamic, exp1,exp2,exp3}.

One significant challenge in employing ansatz for VQLS is that these parameterized quantum circuits with a certain circuit depth do not inherently span the entire space of the unitary matrices \cite{exp1}. Although increasing the circuit depth can enhance the expressiveness of the ansatz, it leads to longer optimization times of the circuit's parameters and the potential emergence of `barren plateaus,’ where gradients vanish during optimization, making it challenging \cite{BP1,BP2,BP3,BP4,BP5}. Consequently, there is no guarantee that the ansatz will always converge to the optimal solution and its performance can be limited by these factors.

To address this limitation, we introduce a novel approach in this study, wherein we incorporate preconditioning strategies as a crucial step in VQLS. Preconditioning, a technique commonly used in iterative Krylov subspace methods, aims to modify a system to improve its numerical properties \cite{Krylov1, Krylov2,Krylov3,Krylov4}. This tailored modification results in a matrix with enhanced spectral properties or a closer approximation to a diagonal matrix, leading to improved convergence.

In our study, we aim to leverage this aspect of preconditioning to reduce the required circuit depth for the ansatz in VQLS. Specifically, we focused on employing incomplete LU (ILU) factorization as a preconditioning method for linear equations \cite{ILU1,ILU2,ILU3,ILU4,ILU5}. ILU factorization is a technique that approximates the LU factorization of a matrix, which is the process of factorizing a matrix into a lower triangular matrix (L) and an upper triangular matrix (U). This approximation is particularly useful in sparse matrix scenarios, where it reduces computational complexity while capturing the essential features of the matrix. By enhancing the spectral properties of the matrix involved in the linear equations with ILU factorization, we hypothesize that the depth of the circuit necessary for the ansatz can be reduced. This potential reduction in the circuit depth could lead to more efficient implementations of VQLS, making it a practical and powerful tool for applications in NISQ computing. 

\section{Preconditioning with ILU factorization}
Here, we discuss the use of ILU factorization as a preconditioning technique for classical iterative Krylov subspace methods. ILU plays a crucial  role in solving linear equations of the form $A\textbf{x} = \textbf{b}$, where $A$ is a given matrix; \textbf{x} and \textbf{b} are vectors. The essence of ILU lies in approximating the exact LU factorization of matrix $A$. In LU factorization, matrix $A$ is factored into two matrices, $L$ and $U$, where $L$ is a lower triangular matrix and $U$ is an upper triangular matrix, and the factorization can be expressed as $A = LU$.

However, in the case of ILU, the goal is to create an approximation of $L$ and $U$ that is computationally less expensive and particularly beneficial for large sparse matrices. The ILU factorization is represented as
\begin{equation}
    A \approx \tilde{L}\tilde{U},
\end{equation}
where $\tilde{L}$ and $\tilde{U}$ are the incomplete lower and upper triangular matrices, respectively. The approximation involves dropping certain elements in $L$ and $U$, typically based on a threshold or  pattern-based strategy, to maintain sparsity and manage computational complexity.

In the context of sparse matrices, ILU factorization aims to approximate LU factorization by filling only the positions in $L$ and $U$, where $A$ has non-zero elements. This was performed to control computational complexity by focusing only on the significant elements of $A$. This can be expressed using the set notation as follows: Let $\Omega$ be the set of indices where $A$ has nonzero elements. Then, the elements of $\tilde{L}$ and $\tilde{U}$ are filled only for those indices in $\Omega$. Mathematically, this can be represented as
\begin{equation}
    \tilde{l}_{ij}, \tilde{u}_{ij} \neq 0 \quad \text{only if} \quad (i, j) \in \Omega,
\end{equation}
where $\tilde{l}_{ij}$ and $\tilde{u}_{ij}$ are the elements in the $i$-th row and $j$-th column of matrices $\tilde{L}$ and $\tilde{U}$, respectively.
To apply this to preconditioning, we consider the product $\tilde{L}\tilde{U}$ as the preconditioning matrix $M$, and its inverse is applied to the left side of the linear equation. The preconditioned system using ILU can then be expressed as
\begin{equation}
    \label{pre}
    M^{-1}A\mathbf{x} = M^{-1}\mathbf{b},
\end{equation}
where $M = \tilde{L}\tilde{U}$. This transformation of the system aims to improve the condition number of matrix $A$, thereby enhancing the convergence properties of the classical iterative methods used for solving the linear equation.

\section{Preconditioning for VQLS}
In this section, we discuss the theoretical underpinnings of the VQLS and explore the application of preconditioning techniques. Our focus is on elucidating the fundamental principles governing the VQLS and examining how classical preconditioning methods can be integrated to enhance the efficiency of the algorithm in solving linear equations.

VQLS is a quantum algorithm designed to solve linear equations of the form $A\textbf{x} = \textbf{b}$, where $A$ is a given matrix and $\textbf{x}$ and $\textbf{b}$ are vectors. The goal is to determine the vector $\textbf{x}$ that satisfies this equation.

The key idea behind VQLS is to represent the solution $\textbf{x}$ as a quantum state $\ket{\textbf{x}}$ and minimize a cost function that quantifies the difference between $A\ket{\textbf{x}}$ and $\ket{\textbf{x}}$. The cost function is defined as
\begin{equation}
    \label{cost}
    C = 1 - \frac{|\braket{\mathbf{b}|A|\mathbf{x}}|^2}{\braket{\mathbf{b}|\mathbf{b}} \braket{\mathbf{x}|A^\dagger A|\mathbf{x}}}.
\end{equation}
This cost function is designed to be minimized when 
$A\ket{\textbf{x}}$ is proportional to $\ket{\textbf{b}}$. The VQLS algorithm iteratively adjusts the parameters of a variational quantum circuit to prepare the state 
$\ket{\textbf{x}}$ such that it minimizes the cost function $C$.

The variational quantum circuit used in VQLS is parameterized by a set of angles $\theta$, and state $\ket{\textbf{x}}$ is prepared as $\ket{\textbf{x}(\theta)}$. Initially, $\ket{\textbf{x}(\theta)}$ does not necessarily represent the solution to the linear equation; however, it is iteratively adjusted through the classical optimization process to approximate the solution of $A\mathbf{x} = \mathbf{b}$. This transformation is expressed as follows:
\begin{equation}
    \ket{\mathbf{x}(\theta)} = V(\theta)\ket{\mathbf{0}}.
\end{equation}
Similarly, the unitary operation $W$ is used to transform the initial state $\ket{\mathbf{0}}$ into state $\ket{\textbf{b}}$, which represents the known vector $\textbf{b}$ in the linear equation. This transformation is essential for evaluating the cost function and is represented as
\begin{equation}
    \frac{\ket{\mathbf{b}}}{\sqrt{\braket{\mathbf{b}|\mathbf{b}}}} = W\ket{\mathbf{0}}.
\end{equation}
The efficient implementation of unitary operations $A$, $V(\theta)$, and $W$ is crucial in the VQLS. If $A$ is Hermitian, matrix $A$ can be decomposed into a sum of unitary matrices as $A = \sum_{k} \alpha_{k} A_{k}$. Even if $A$ is not Hermitian, it can be Hermitianized by either setting $A'=A^\dagger A$ or using an ancillary qubit to construct $A'=\left(\begin{array}{rr}0 & A \\ A^\dagger & 0 \\ \end{array}\right)$. This decomposition enables the embedding of $A$ into a quantum circuit. Notably, the efficiency of this embedding is significantly enhanced when $A$ is a sparse matrix. Sparse matrices allow for a more streamlined and resource-efficient implementation of the unitary matrices $A_k$ in the quantum circuit.

Furthermore, it is typically challenging to efficiently embed classical data $\textbf{b}$ as a quantum state $\ket{\textbf{b}}$ using $W$ \cite{preparation}. Several techniques have been proposed in recent years \cite{embed1,embed2,embed3,embed4,embed5,embed6}, primarily within the field of quantum machine learning \cite{embed7,embed8}.

In our approach to optimizing \( V(\theta) \) in VQLS, we focus on reducing the necessary circuit depth of the ansatz through classical preconditioning. Instead of directly solving the linear system \( A\mathbf{x} = \mathbf{b} \), we solve the preconditioned system defined by Eq. (\ref{pre}). In this transformation, the original matrix \( A \) and vector \( \mathbf{b} \) are preconditioned using ILU factorization, resulting in a new system \( \tilde{A}\mathbf{x} = \tilde{\mathbf{b}} \), where \( \tilde{A} = M^{-1}A \) and \( \tilde{\mathbf{b}} = M^{-1}\mathbf{b} \).

A typical ansatz is shown in Fig. \ref{fig1}(a), known as a hardware-efficient ansatz corresponding to $V(\theta)$ in conventional VQLS. This ansatz structure comprises local phase rotation gates with independently adjustable parameters and CNOT gates between adjacent qubits, repeated for $D$ cycles.

Fig. \ref{fig1}(b) illustrates the effect of preconditioning on reducing the required depth $D$ of the ansatz in VQLS. By applying preconditioning, the space of quantum states that needs to be explored by the ansatz can be narrowed, potentially reducing the necessary circuit depth .

Fig. \ref{fig1}(c) depicts the impact of preconditioning on the search space in the Krylov subspace method for classical iterative linear solvers. Preconditioning can help to focus the iterative search within a smaller subspace, thereby improving the efficiency of the classical solver.

To reduce the search space required, as shown in Fig. \ref{fig1}(b), the $V(\theta)$ required for obtaining the optimal solution must evolve from `complex' to `simpler' by preconditioning. In conventional VQLS, $V(\theta)$ is typically defined as the operator that transforms state $\ket{\mathbf{0}}$ into $\ket{\mathbf{x}(\theta)}$. However, in our approach, we redefine $V(\theta)$ as the operator that transforms the preconditioned state $\ket{\tilde{\mathbf{b}}}$ into $\ket{\mathbf{x}(\theta)}$ as
\begin{equation}
    \ket{\mathbf{x}(\theta)} = V(\theta)\tilde{W}\ket{\mathbf{0}},
\end{equation}
where $\tilde{W}\ket{\textbf{0}} = \ket{\tilde{\textbf{b}}}/\sqrt{\braket{\tilde{\textbf{b}}|\tilde{\textbf{b}}}}$.

Based on this definition, the cost function in Eq. (\ref{cost}) can be rewritten as
\begin{equation}
    \label{costf}
    C = 1 - \frac{|\braket{\tilde{\textbf{b}}|\tilde{A}V(\theta)|\tilde{\textbf{b}}}|^2}{\braket{\tilde{\textbf{b}}|\tilde{\textbf{b}}} \braket{\mathbf{x}(\theta)|\tilde{A}^\dagger \tilde{A}|\mathbf{x}(\theta)}}.
\end{equation}
$\tilde{A}$ can be embedded into a quantum circuit as a sum of unitary matrices, expressed as $\tilde{A} = \sum_{k} \tilde{\alpha}_{k} \tilde{A}_{k}$. Following this formulation and Eq. (\ref{costf}), the cost function $C$ can be computed on a quantum computer using the Hadamard tests for $\bra{\tilde{\mathbf{b}}}\tilde{A}_{k}V(\theta)\ket{\tilde{\mathbf{b}}}$ and for $\bra{\tilde{\mathbf{x}}(\theta)}\tilde{A}_{k'}\tilde{A}_{k}\ket{\tilde{\mathbf{x}}(\theta)}$. 

In Ref. \cite{vqls1}, the authors proposed a definition of the cost function given by local measurements to mitigate the barren plateau. However, in this study, we focused on exploring the ability of preconditioning to reduce the necessary depth of the ansatz. Therefore, for numerical demonstrations , we adopted the cost function defined in Eq. (\ref{costf}).

\begin{figure}[H]
    \includegraphics[width=8.4cm]{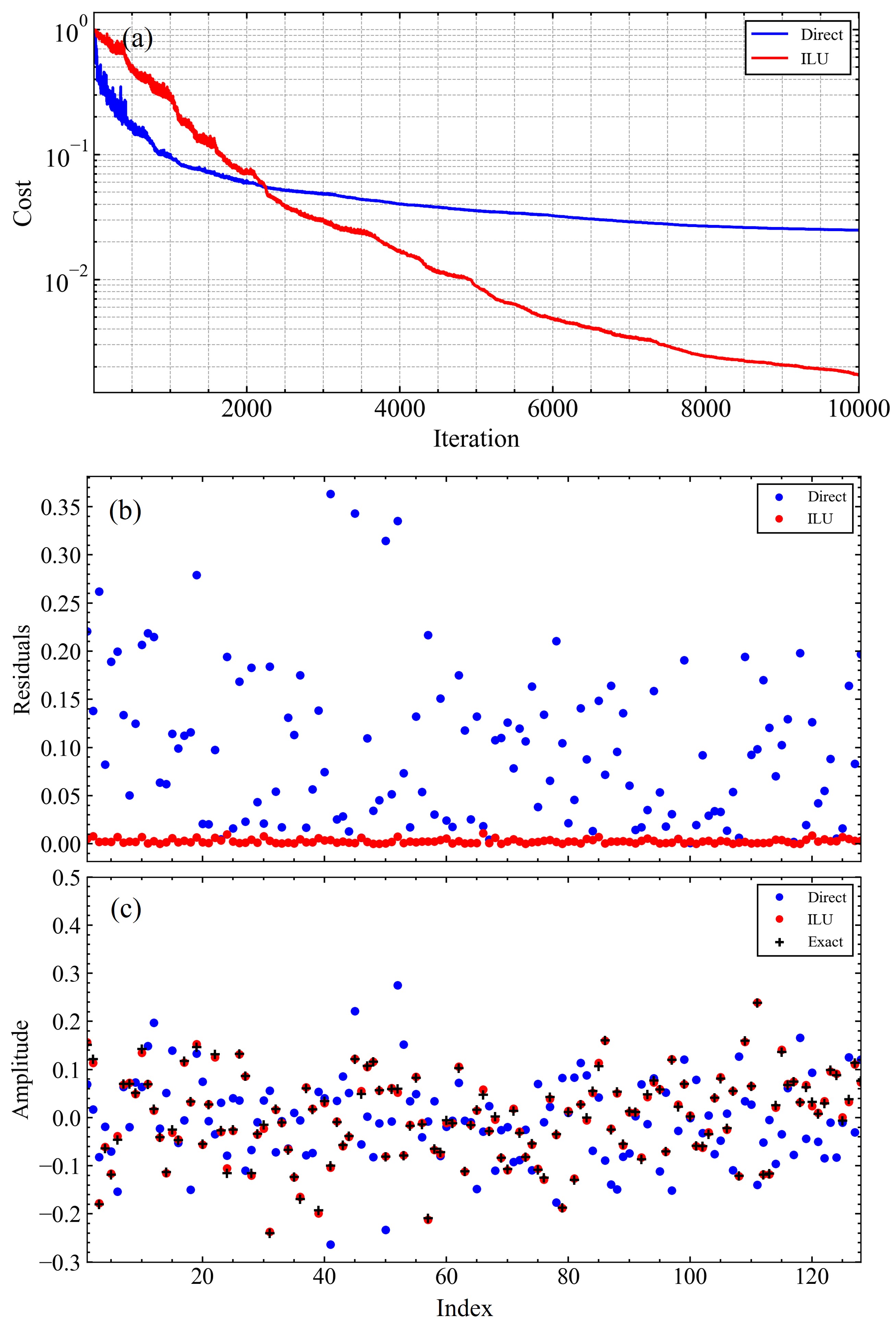} 
    \caption{(a) A graph comparing the results of solving linear equations by VQLS without preconditioning (blue line) versus with preconditioning (red line). (b) A graph showing the absolute values of residuals between the obtained solution and the exact solution calculated using matrix inversion. (c) A plot showing the obtained solution from VQLS and the exact solution.}
    \label{fig2}
\end{figure}

\section{Numerical demonstration }
We applied ILU factorization as a preconditioning to a linear equation defined by a \(128 \times 128\) sparse matrix and numerically  investigated the effects of this preconditioning. A variational quantum circuit of this size can be described using an 8-qubit quantum circuit because seven qubits are used to represent the \(2^7 = 128\) Hilbert space, and an ancilla qubit is used to convert to the Hermite matrix. In the tested linear equation, both the matrix \(A\) and vector $\mathbf{b}$ comprised entirely real numbers; thus, we adopted a hardware-efficient ansatz consisting of RY and CNOT gates, as depicted in Fig. \ref{fig1}(a). This configuration ensures that all amplitudes remain real. The density of matrix \(A\) was set to 0.2, and the elements of \(A\) and \(b\) were generated using uniform random numbers ranging from $-1$ to $1$. The number of iterations was set to 10,000, and the Adam optimizer with a learning rate of 0.001 was used to optimize  the ansatz parameters.

The results demonstrating the effects of preconditioning under the condition of a circuit depth of 20 are presented in Fig. \ref{fig2}. After 10,000 iterations, the VQLS with preconditioning exhibited improved cost convergence and residuals compared to those without preconditioning, with respect to the exact solution. The results presented in Fig. \ref{fig2} pertain to a particular involving a randomly generated sparse matrix. We observed similar improvements in solution convergence owing to  the preconditioning of the other linear equations generated using the same procedure.

\begin{figure}[H]
    \includegraphics[width=8.4cm]{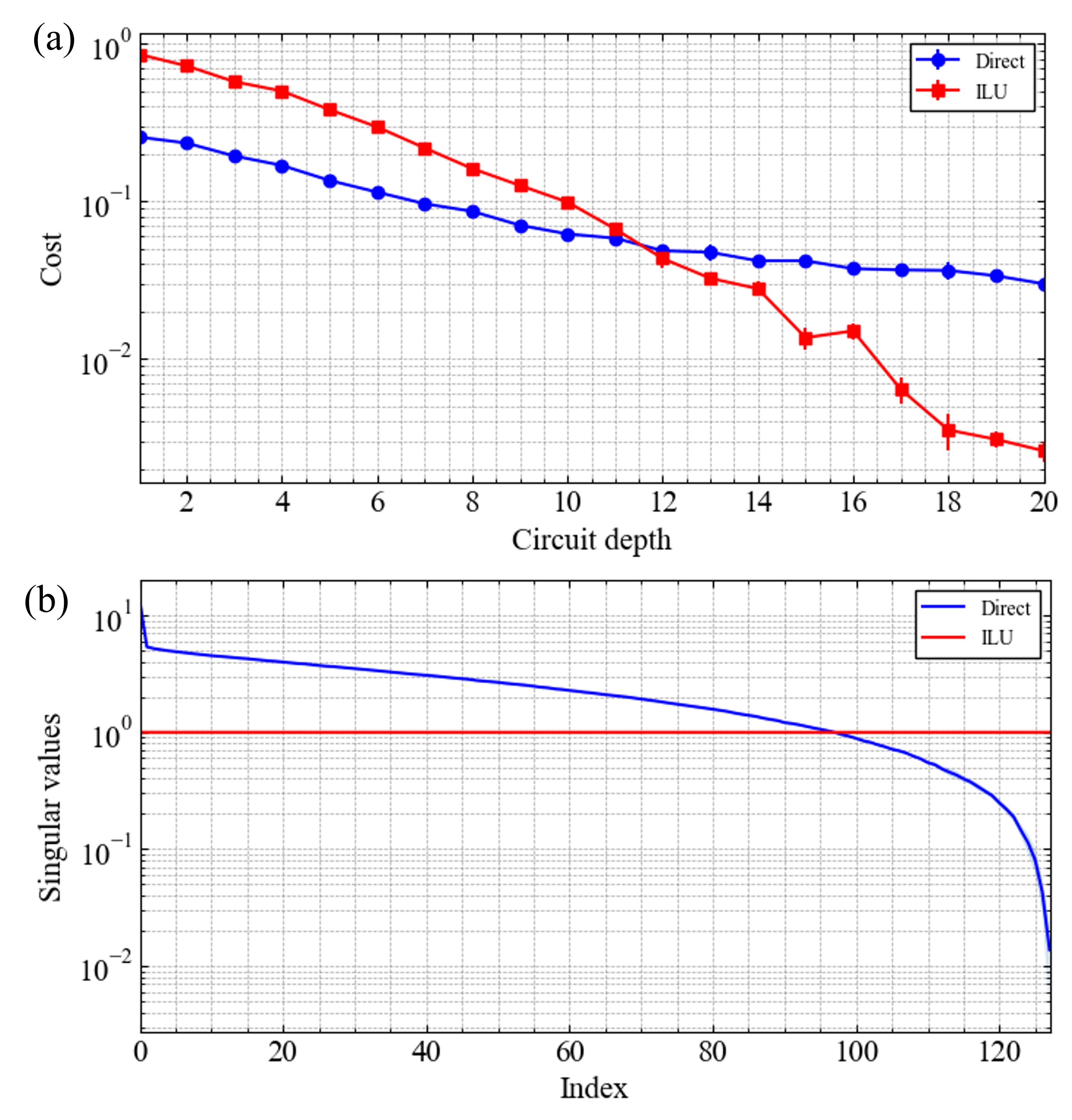} 
    \caption{(a) A plot of the cost obtained after iterations at each circuit depth. The blue line represents the results when VQLS was directly employed, and the red line shows the results with preconditioning applied. It displays the average values and standard errors for 10 random linear equation cases. (b) A comparison of the singular value distributions before (blue line) and after (red line) preconditioning for 10 random sparse matrices. The standard error was negligible on the graph.}
    \label{fig3}
\end{figure}

To demonstrate the reduction in the required depth of the ansatz owing to preconditioning, Fig. \ref{fig3}(a) presents the average cost obtained after 10,000 iterations for linear equations defined by 10 different  \(128 \times 128\) random sparse matrices, with the ansatz's circuit depth varying from 1 to 20. As shown in the figure, preconditioning reduces the complexity of the unitary operations that the ansatz must represent, leading to an improved convergence of the solution with respect to the circuit depth.

To investigate the reason for the improved solution convergence, the average spectra of the singular values before and after preconditioning for the 10 random sparse matrices are presented in Fig. 3(b). Although the original matrices exhibited a skew in singular values, preconditioning led to a uniform distribution in the spectrum. This implies that the preconditioning brought matrix $A$ closer to the identity matrix, enabling the ansatz to generate solutions with less entanglement.
\\

\begin{figure}[b]
    \includegraphics[width=8.4cm]{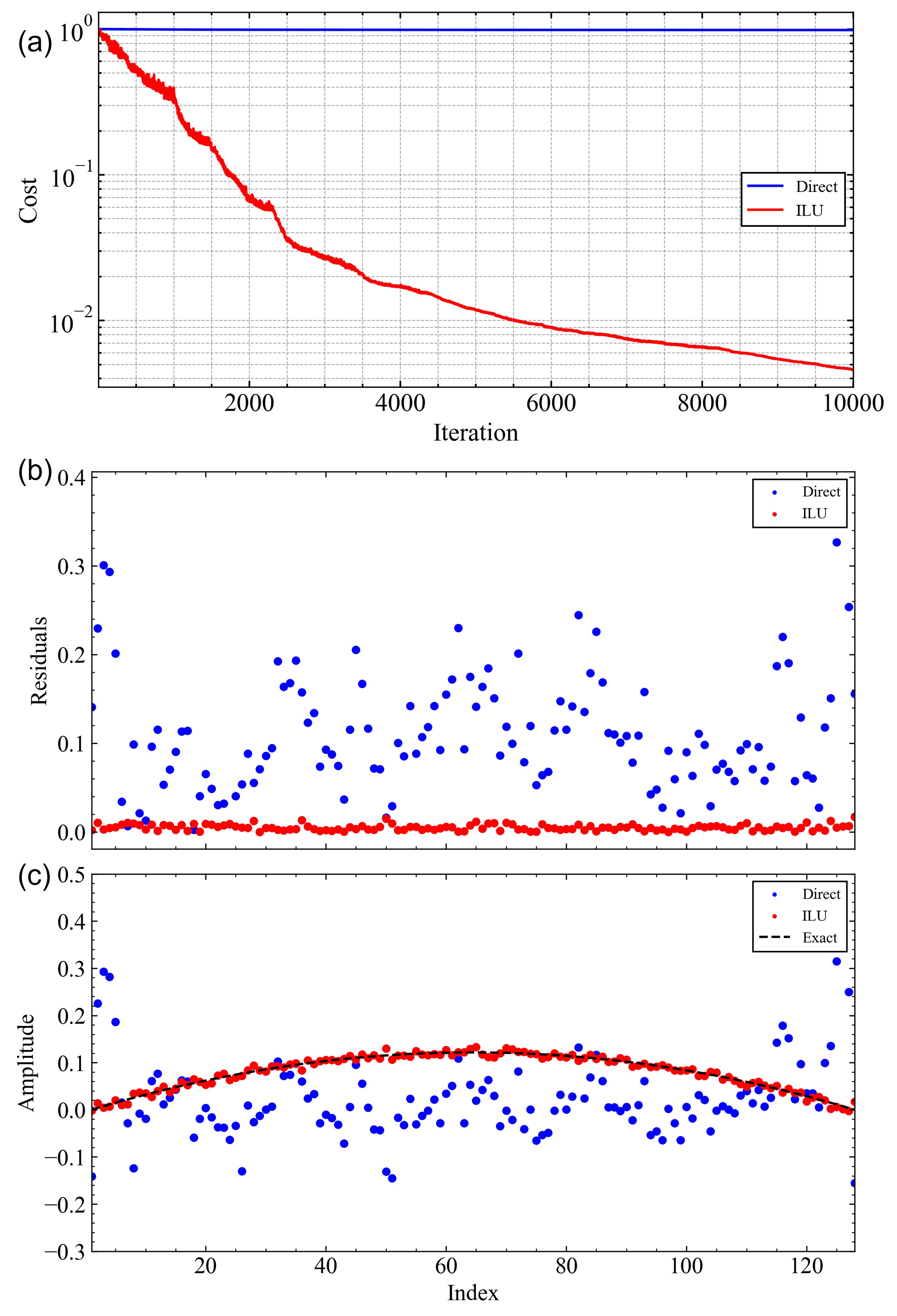} 
    \caption{(a) A graph comparing the results of solving the diffusion equation by VQLS without preconditioning (blue line) versus with preconditioning (red line). (b) A graph showing the absolute values of residuals between the obtained solution and the exact solution calculated using matrix inversion. (c) A plot showing the obtained solution from VQLS and the exact solution.}
    \label{fig4}
\end{figure}

\section{Application to Real-World Problems }
The solution of linear equations is a crucial task that appears in real-world scenarios. Here, as an example of the linear equations that emerge in real-world contexts, we solve the steady-state solution of a one-dimensional heat diffusion equation.

We benchmark the effectiveness of the preconditioned VQLS by solving the simplest example of heat diffusion found in standard textbooks. Both ends of the rod are subject to boundary conditions with constant temperatures, and assuming that the conductor uniformly generates heat at a constant rate $f$, the differential equation representing this problem is given as follows:

\begin{equation}
\left\{ \,
    \begin{aligned}
    & -\frac{d^2u}{dx^2} = f \\
    & u(0) = 0, u(L) = 0
    \end{aligned}
\right.
\end{equation}

By discretizing this equation, it can be converted into a linear equation. Considering $\Delta x$ is sufficiently small in the interval $i=2$ to $i=N-1$, we have $-(u_{i-1}-2u_i+u_{i+1})/\Delta h^2 = f$. Furthermore, the boundary conditions are $u_0=0$ and $u_N=0$.
When $\Delta h^2$. From these relations, the coefficients of the linear equation $A\bf{x}=\bf{b}$ can be determined.

Figure \ref{fig4} illustrates the application of preconditioning to the solution of the steady-state heat- diffusion equation. In this example, we have discretized the 1D  rod into $128$ segments. The linear equation resulting from the discretization of the differential equation was solved using VQLS, both with and without preconditioning.

Figure \ref{fig4}(a) shows the convergence of the cost function in both cases. The blue line represents the cost function without preconditioning, whereas the red line represents the cost function with preconditioning. Notably, in this example, vector $b$ contains only low-frequency components, as opposed to a more general case in which $b$ can contain a mixture of low- and high-frequency components. In such scenarios, where $b$ is dominated by low-frequency components, the convergence of the VQLS algorithm without preconditioning can become particularly challenging. This is because ansatz might struggle to efficiently represent solutions that are predominantly low frequency, leading to slower convergence rates. Conversely, the application of preconditioning can significantly improve the convergence by effectively transforming the problem into a form that is easier for the ansatz to represent, thereby enhancing the overall performance of the VQLS algorithm.

Figure \ref{fig4}(b) shows the absolute values of the residuals between the solution obtained from VQLS and the exact solution calculated using classical matrix inversion techniques. The residuals are lower for the preconditioned case (red line) than for the non-preconditioned case (blue line), demonstrating that preconditioning accelerates the convergence and enhances the accuracy of the solution obtained from VQLS.

Finally, Figure \ref{fig4}(c) presents a comparison between the solution obtained from the VQLS (both with and without preconditioning) and the exact solution. The blue dots represent the solution obtained without preconditioning, the red dots represent the solution obtained with preconditioning, and the black line represents the exact solution. The plot illustrates that the solution obtained with preconditioning is closer to the exact solution, further validating the effectiveness of preconditioning in improving the performance of the VQLS for real-world problems such as the heat diffusion equation.

\section{Discussion }
Preconditioning using classical computers can reduce the amount of entanglement required for the VQLS. However,  this preconditioning might increase the amount of entanglement needed to generate $\ket{\mathbf{b}}$. Typically, generating any arbitrary $\ket{\mathbf{b}}$ requires $O(N^2)$ gates \cite{preparation}; however, for some linear equations, $\ket{\mathbf{b}}$ can be represented with fewer gates. However, the application of preconditioning may complicate the problems that were previously advantageous for efficiently generating $\ket{\mathbf{b}}$, potentially increasing the number of gates required to generate $\ket{\tilde{\mathbf{b}}}$.

However, adding preconditioning could potentially reduce the number of gates required to generate $\ket{\mathbf{b}}$, and the best strategy is to choose a preconditioner that improves the properties of $A$ while reducing the number of gates required to generate $\ket{\mathbf{b}}$.

In our numerical demonstration, we adopted ILU as the preconditioner,  several methods are known as preconditioners. When choosing a preconditioner, it is necessary to consider the balance between three costs: the number of gates required to generate $\ket{\mathbf{b}}$, the depth of the ansatz required to obtain the solution, and the additional computational cost required for preconditioning on a classical computer.

\section{Conclusion }
We applied preconditioning, commonly used in Krylov subspace methods, to variational quantum circuits and demonstrated its effectiveness numerically. By applying ILU to 10 instances of linear equations given by $128 \times 128$ random sparse matrices, we observed 
a reduction in the required depth of the ansatz in all cases owing to preconditioning.

Reducing the depth of the ansatz is important for decreasing the number of iterations needed for optimization in variational quantum algorithms and crucial for increasing noise resilience and avoiding barren plateaus in NISQ machines. By incorporating classical computing assistance, such as preconditioning, it is conjectured that the calculation costs of quantum computers can be reduced, and more accurate computational results can be obtained with NISQ machines.

\bibliographystyle{apsrev}
\bibliography{main}

\end{document}